\documentclass[preprintnumbers,showpacs,amsmath,amssymb,floatfix,9pt,prd,onecolumn,
superscriptaddress,nofootinbib]{revtex4}
\usepackage{graphicx}
\usepackage{latexsym}
\usepackage{epsfig}
\usepackage{amssymb}
\usepackage[utf8]{inputenc}
\begin{document}

\title{Relativistic compact stars in Tolman spacetime via an anisotropic approach }

\author{Piyali Bhar }
\email{piyalibhar90@gmail.com; piyalibhar@associates.iucaa.in}
\affiliation{Department of
Mathematics,Government General Degree College, Singur, Hooghly, West Bengal 712 409,
India}

\author{Pramit Rej \footnote{Corresponding author}}
\email{pramitrej@gmail.com
 } \affiliation{Department of
Mathematics, Sarat Centenary College, Dhaniakhali, Hooghly, West Bengal 712 302, India}

\author{P. Mafa Takisa}
\email{pmafakisa@gmail.com}
\affiliation{Science, Engineering and Technology, Department of Physics,
University of South Africa, Florida, Johannesburg 1710, South Africa}

\author{M. Zubair}
\email{mzubairkk@gmail.com;drmzubair@cuilahore.edu.pk}
\affiliation{Department of Mathematics, COMSATS University Islamabad, Lahore Campus, Lahore, Pakistan}

\begin{abstract}

In this present work, we have obtained a singularity-free spherically symmetric stellar model with anisotropic pressure in the background of Einstein's general theory of relativity. The Einstein's field equations have been solved by exploiting Tolman {\em ansatz} [Richard C Tolman, Phys. Rev. 55:364, 1939] in $(3+1)$-dimensional space-time. Using observed values of mass and radius of the compact star PSR J1903+327, we have calculated the numerical values of all the constants from the boundary conditions. All the physical characteristics of the proposed model have been discussed both analytically and graphically. The new exact solution satisfies all the physical criteria for a realistic compact star.
The matter variables are regular and well behaved throughout the stellar structure.
Constraints on model parameters have been obtained. All the energy conditions are verified with the help of graphical representation. The stability condition of the present model has been described through different testings.
\end{abstract}



\maketitle

\section{Introduction}

Stellar evolution predicts that when the nuclear fuel gets exhausted, the stars turn into highly dense compact objects such as white dwarf, neutron star or back hole. Massive stars undergoing the supernova explosion turn into neutron star and black hole. For neutron star, the main idea is that the gravitational collapse is supported by the neutron degeneracy pressure. The general perception is that for high densities at the core, nucleons have to converted to hyperons or either form condensates. Some studies predict that these nucleons could form Cooper pairs and can be in superfluid state. Based on the MIT bag model, Witten \cite{Witten1984} provides the existence of strange quark matter, which indicates that the quarks inside the compact objects might not be in a confined hadronic state. At the high densities and pressures they could form a larger colorless region with equal part of up, down and strange quarks. Consequently, the composition of the core region of compact objects is still an open subject in relativistic astrophysics.\par

When densities of compact stars are greater than the nuclear matter density, it expects the appearance of unequal principal stresses, called anisotropic effect. This usually means that the radial pressure component $p_r$ is not equal to the transverse component $p_t$. The presence of anisotropy was first predicted for self-gravitating objects in Newtonian regime by Jeans \cite{Jeans1922}. Later, Lemaitre \cite{Lemaitre1933} considered the local anisotropy effect in the context of general relativity and showed that the presence of anisotropy can change the upper limits on the maximum value of the surface gravitational potential.
Ruderman \cite{Ruderman1972} showed that a compact star with matter density $(\rho > 10^{15} \mbox{g cm}^{-3})$,
where the nuclear interaction become relativistic in nature, is likely to be anisotropic.
Herrera \cite{Herrera1997} presented the evidence on the appearance of local anisotropy in self gravitating systems in both Newtonian and general relativistic context. Since then, a lot of investigations have been carried out in finding new exact solutions with anisotropy feature.\par
For half of century, the theory of anisotropic compact stars in General Relativity has been developed.
Bower and liang \cite{Bowers1974} provided the generalization of Tolman-Oppenheimer-Volkov equation in presence of anisotropy. The stability of a stellar model can be enhanced by a presence of a repulsive anisotropic force
when $ \Delta=p_t - p_r > 0 $. This feature leads to more compact stable configurations compare to the
isotropic case as pointed out by Ivanov \cite{Ivanov2002}. The work of Ivanov \cite{Ivanov2002} gives general bounds on the redshift for any anisotropic compact objects. Cosenza et al. \cite{cosenza} developed a heuristic way for the modeling of stars with anisotropic fluid distribution. Herrera et
al. \cite{herrera2004} formulated governing equations with anisotropic stress for self-gravitating spherically symmetric distributions. Herrera and Barreto \cite{herrera2013,herrera2013a} described a new way to study
stability of polytropic models by means of Tolman-mass. Errehymy and Daoud \cite{erreh} obtained analytical solution using
dark-energy (DE), which is characterized by a equation of
state (EoS) of the type $p_r = (\gamma-1)\rho$. Singh et al. \cite{singh2020a} presented the anisotropic stars by taking a modified polytropic equation of
state in the framework of the Korkina-Orlyanskii
spacetime. In the background of the conformal motion and under Karmarkar condition \cite{karmarkar1948gravitational}, several researchers studied the model of compact stars in presence of anisotropy which can be found in refs. \cite{Errehymy:2021cdl,Ramos:2021drk,Zubair:2021uek,Mustafa:2021mgh,Mustafa:2021aci,Bhar:2017hbw,Bhar:2017ynp,Rahaman:2010dg,Bhar:2014jta}. The anisotropic compact star models have also been studied in the context of modified gravity. Zubair and Abbas \cite{zubair2016} explored charged anisotropic compact stars in $f(T)$ gravity based on the diagonal form of tetrad field for static spacetime. Biswas and his collaborators \cite{biswas2019} used metric potentials given by Krori-Barua \cite{kb75} and established a new model for highly compact anisotropic strange system in the context of $f(R,\,T)$ gravity. Zubair et al \cite{zubair2021} obtained anisotropic compact star models in $f(T)$ gravity
with Tolman-Kuchowicz spacetime. Bhar \cite{bhar2020} explored a charged model of compact star in $f(R,\,T)$ gravity admitting chaplygin equation of state. Rej and Bhar \cite{rej2021a} obtained charged strange star in $f (R,\,T )$ gravity with linear equation of state. In the framework of Teleparallel Gravity, Nashed et al. \cite{Nashed:2020kjh} derived a charged non-vacuum solution for a physically symmetric tetrad field with two unknown functions of radial coordinate. Singh et al. \cite{Singh:2020iqh} explored exact solutions free from any physical and geometrical singularities, as well as the existence of compact stellar systems throughout linear and Starobinsky $f(R,T)$ gravity theory. More details about the anisotropic models of compact stars can be found in some of these Refs. \cite{Herrera2008,Thirukkanesh2008,Varela2010,Bhar2015a,Bhar2015b,Singh2016,Maurya2016,Mafa2017,Estevez-Delgado2018,Das2019,Bhar2019,Mafa2019}.\par

Now-a-days it is a concern of famed interest to obtain the models of early mentioned small sized heavily dense stellar objects to advocate the regime of strictly coupled gravitational fields, since long time, general relativity (GR) theory is a handy tool to understand the behavior of these heavily compact objects. With considerable observationally and experimentally subsidized support, Will \cite{Will2005} explained its gravitational consequences and interactions in four dimensional metric-space in admissible manners. Compact stars are normally understood as spherically symmetric and isotropic highly-dense objects. However anisotropy favors the heterogeneous pressure conditions and generalizes the isotropic conditions. In this study we use the Tolman spacetime \cite{Tolman1939} as spherically symmetry to explore the anisotropy of the heavenly objects named as compact stars. Several authors mentioned in the refs. \cite{Bhar2016,Zubair2020,Shee2018,Ray2020,Singh2020} used this spacetime to investigate the structures of compact stars in their research articles. Tolman-Kuchowiz spacetime \cite{Tolman1939,Kuchowicz1968} have already been used by the authors \cite{Jasim2018,Maurya2019,Bhar2019a,Shamir2020} to discuss the anisotropic manners of fluid in compact star formation.\par

In this work, we have investigated the physical stability and viability of anisotropic model of compact stars in the background of the
Tolman spacetime. We use the physical requirements for acceptability of the model to imply constraints on the maximum allowed compactness and redshift. We further use pulsars observational data of total mass and radius to test the model validity under a certain boundary conditions. We have considered the following compact stars whose observational mass and radius are given in the bracket: Vela X -1 [mass=($1.77 \pm 0.08)M_{\odot}$; radius=$9.56_{-0.08}^{+0.08}$ km] \cite{Rawls:2011jw}, LMC X -4 [Mass=($1.04 \pm 0.09)M_{\odot}$; radius= $8.301_{-0.2}^{+0.2}$ km] \cite{Rawls:2011jw}, 4U $1608 - 52$ [mass=($1.74 \pm 0.14)M_{\odot}$; radius= $9.528_{-0.15}^{+0.15}$ km] \cite{Guver:2010td}, PSR J$1614 - 2230$ [mass=$(1.97 \pm 0.04)M_{\odot}$; radius= $9.69_{-0.2}^{+0.2}$ km] \cite{Demorest:2010bx},  EXO $1785 - 248$ [mass= ($1.3 \pm 0.2) M_{\odot}$; radius=$8.849_{-0.4}^{+0.4}$] \cite{Ozel:2008kb} and from our present work, we have successfully estimated their masses and radii.\par
The outline of our paper is as follows: In section \ref{sec2} we have described the basic field equations and their solutions. We picked the Tolman ansatz for one metric coefficient $W=1+ar^2+br^4$ and calculated the relation of other components by employing the role of anisotropy factor. Some basic physical properties of the model are also discussed including the density, pressure, energy conditions and mass-radius relation. Section \ref{sec3} deals with the exterior spacetime and smooth matching conditions. We have calculated unknown constants using the matching conditions and results are given in Table \ref{table1}. The relationship between the matter density and pressure have been discussed in section \ref{sec4}. It is worthwhile to mention that we here obtain the nonlinear EoS which contains the attributes of modified Chaplygin gas. The next section gives an idea about the stability and equilibrium condition of our model under different forces. Here, we probe the viability of our results under the Harrison-Zeldovich-Novikov stability criterion, causality condition and cracking, realistic adiabatic index and TOV equation.
The final section \ref{sec6} briefly summarizes this manuscript and highlights the major conclusions drawn.

\section{Interior Spacetime}\label{sec2}
\subsection{Basic field Equations}
To describe the interior of a compact star model, let us assume that the matter within the star is anisotropic in nature and therefore, the corresponding energy-momentum tensor is given as,
\begin{equation}\label{1h}
T_{\nu}^{\mu}=(\rho c^2+p_t)~\xi^{\mu}\xi_{\nu}-p_t g_{\nu}^{\mu}+(p_r-p_t)~\zeta^{\mu}\zeta_{\nu},
\end{equation}
with $ \xi^{i}\xi_{j} =-\zeta^{i}\zeta_j = 1 $ and $\xi^{i}\zeta_j= 0$. Here the vector $\zeta^{i}$ is the space-like vector and $\xi_i$ is the fluid 4-velocity and which is orthogonal to $\zeta^{i}$, $\rho$ is the matter density, $p_t$ and $p_r$ are respectively the transverse and radial pressure components of the fluid and these two pressures act orthogonally to each other.\\
In (3+1)-dimension, in Schwarzschild coordinates
$x^{\mu} = (t,\,r,\,\theta,\,\phi)$, a static and spherically symmetry spacetime  is described by the line element,
\begin{equation}\label{2}
ds^{2}=-V^{2}dt^{2}+W^{2} dr^{2}+r^{2}(d\theta^{2}+\sin^{2}\theta d\phi^{2}),
\end{equation}
where the metrics $V$ and $W$ are functions of the radial coordinate `r' and do not depend on the time `t' i.e., they are static. The asymptotic flatness implies $\lim_{r\rightarrow \infty}V(r)~=0=~\lim_{r\rightarrow \infty}W(r)$ and the regularity at the center imposes the condition
$V(0)=$ constant and $W(0)=1$ \cite{Banerjee:2020dad}.\par
The Einstein field equations for our present anisotropic model of compact star using the energy momentum tensor (\ref{1h}) are described by,
\begin{eqnarray}\label{d8}
G_{\nu}^{\mu}&=&R_{\nu}^{\mu}-\frac{1}{2}g_{\nu}^{\mu}R=\frac{8\pi G}{c^4} T_{\nu}^{\mu},
\end{eqnarray}
Here $G_{\nu}^{\mu}$ is the Einstein tensor, $G$ and $c$ are respectively the universal gravitational constant and speed of the light.\par
Using (\ref{d8}) and assuming $G=1=c$, the Einstein field equations can be expressed as the following system of ordinary
differential equations:
\begin{eqnarray}
\kappa \rho&=&\frac{1}{r^2}\left(1-\frac{1}{W^2}\right)+\frac{2W'}{W^3r},\label{1a}\\
\kappa p_r&=&\frac{1}{W^2}\left(\frac{1}{r^2}+\frac{2V'}{Vr}\right)-\frac{1}{r^2},\label{2a}\\
\kappa p_t&=&\frac{V''}{VW^2}-\frac{V'W'}{VW^3}+\frac{1}{W^3rV}(V'W-W'V),\label{3a}
\end{eqnarray}
where $\kappa=8\pi$ and `prime' denotes `r' derivative. \\
The mass function, $m(r)$, within the radius `$r$' is introduced by,
\begin{eqnarray}\label{d4}
\frac{1}{W^2}&=&1-\frac{2m(r)}{r},
\end{eqnarray}
Using (\ref{d4}), from (\ref{1a})-(\ref{3a}) one can get,
\begin{eqnarray}
m(r)&=&4\pi\int_0^{r}\omega^{2}\rho(\omega)d\omega,\label{4}\\
\frac{2V'}{V}&=&\frac{\kappa r p_r+\frac{2m}{r^2}}{1-\frac{2m}{r}},\label{k9}\\
\frac{dp_r}{dr}&=&-(\rho+p_r)\frac{V'}{V}+\frac{2}{r}(p_t-p_r),\label{k10}
\end{eqnarray}
Combining (\ref{k9}) and (\ref{k10}), one can finally obtain :
\begin{eqnarray}
\frac{dp_r}{dr}&=&-\frac{\rho+p_r}{2}\frac{\left(\kappa r p_r+\frac{2m}{r^2}\right)}{\left(1-\frac{2m}{r}\right)}+\frac{2}{r}(p_t-p_r).\label{k11}
\end{eqnarray}
The eqn.~(\ref{k11}) is called the Tolman-Oppenheimer-
Volkoff (TOV) equation of a hydrostatic equilibrium
for the anisotropic stellar configuration.\par

Our goal is to generate an exact solution by solving the field equations which does not suffer from any kind of singularities. The following metric provides a singularity free model which will be described in the next sections.
To solve the above set of equations (\ref{1a})-(\ref{3a}), instead of solving the Eqns. for any prescribed equation of state, we are rather interested in solving Eqs. (\ref{1a})-(\ref{3a}) with the help of the following {\em ansatz}
\begin{equation}\label{5}
W^2=1+ar^2+br^4,
\end{equation}
The choice (\ref{5}) was proposed in \cite{Tolman1939},
where `$a,\,b$' are constants of having units length$^{-2}$ and length$^{-4}$ respectively and can be obtained from the boundary conditions which has been discussed in details in section \ref{sec3}. The above metric potential was used earlier by several researchers in the background of General relativity as well as in modified gravity. Jasim et al. \cite{Jasim2018} studied a singularity-free model for
the spherically symmetric anisotropic strange stars under
Einstein's general theory of relativity in presence of the cosmological constant $\Lambda$ which depends on radial co-ordinate $r$. Biswas et al. \cite{Biswas:2019doe} used this metric potential along with the MIT Bag model equation of state to obtain strange star model. Bhar et al. \cite{Bhar:2020tah} successfully obtained a new relativistic compact stellar model by using the above {\em ansatz} in General relativity. Patwardhan and Vaidya \cite{patwardhan1943relativistic}, Mehra \cite{mehra_1966} also used this {\em ansatz} earlier. Singh et al. \cite{singh2016solutions} also used the same metric potential to obtained the compact star model in embedding class I spacetime. On the other hand, we now discuss about the use of this metric potential in the context of modified gravity.
This metric potential was used earlier by Bhar et al. \cite{Bhar2019a} to model compact object in Einstein-Gauss-Bonnet gravity and  Javed et al. \cite{j1} used this metric potential to model anisotropic spheres in
$f (R, G)$ modified gravity,  Biswas et al. \cite{biswas2020} obtained an anisotropic strange star with $f (R, T )$ gravity, Majid and Sharif \cite{majid} obtained
quark stars in massive Brans-Dicke gravity, Naz and Shamir \cite{naz} found the stellar model
in $f (G)$ gravity, Farasat Shamir
and Fayyaz \cite{fs} obtained the model of charged compact star
in $f (R)$ gravity, Rej et al. \cite{rej2021} studied the charged compact star in the context of $f(R,\,T)$ gravity. Since one can obtain the results in the background of General relativity as a limiting case of the result in different modified theories of gravity, one can conclude that all the obtained results in modified gravity are also consistent in the background of General relativity. One can note that this metric potential successfully produces the model of compact star which is singularity free and satisfies all the requirements to be physically acceptable. Therefore, inspired by all the previous works done mentioned earlier in General relativity as well as in modified gravity, we choose the metric potential in equation (\ref{5}) in our present paper.  \par
Using equations (\ref{2a})-(\ref{3a}) and the definition of anisotropy $\Delta=p_t-p_r$, we obtain the following
expression
\begin{eqnarray}
\label{fonc}
\kappa\Delta &=&\frac{1}{W^2}\left(\frac{V''}{V}-\frac{V'}{Vr}\right)-\frac{W'}{W^3}\left(\frac{V'}{V}-\frac{1}{r} \right)
- \frac{1}{W^2 r^2} +\frac{1}{r^2},
\end{eqnarray}
By making following transformation $\frac{1}{W^{2}}=T$,  equation (\ref{fonc}) becomes
\begin{eqnarray}
\label{fonc1}
\kappa\Delta-\frac{1}{r^2} &=&   T\left(\frac{V''}{V}-\frac{V'}{Vr}- \frac{1}{ r^2}\right)+\frac{T'}{2}\left(\frac{V'}{V}+\frac{1}{r}\right).
\end{eqnarray}
The above equation (\ref{fonc1}) can also be written as,
\begin{eqnarray}
\label{fonc2}
\frac{V''}{V}+P(r)\frac{V'}{V}+Q(r)=0
\end{eqnarray}
where
\begin{eqnarray*}
\label{fonc3}
P(r)= \frac{T'}{2T}-\frac{1}{r},\,
Q(r)= \frac{T'}{2Tr}-\frac{1}{r^2}-\frac{\chi}{T},\,
\chi=\kappa \Delta-\frac{1}{r^2}.
\end{eqnarray*}
Eq.~(\ref{fonc2}) is a second order ordinary differential equation (ODE) in $V$. An algorithm was presented by Herrera et al. \cite{Herrera2008} that shows that all static spherically symmetric anisotropic solutions of Einstein's field equations may be generated from Eq. (\ref{fonc2}) by
two generating functions $\Delta$ and $V$. If one can obtain the metric potential $V$, the other
physical variables may be expressed in terms of the functions $\chi$ and $V$.
\subsection{Solution of field equations and pressure anisotropy}
From Eq.~(\ref{fonc2}), it is clear that once we assume the anisotropic factor $\Delta$, we can easily solve the second order ODE and consequently obtained the expression for $V$.
The anisotropy $\Delta$ should be chosen in such manner that
\begin{itemize}
  \item it should vanish at the center of the star,
  \item it does not suffer from any kind of singularities,
  \item $\Delta$ is positive inside the stellar interior and finally
  \item the field equation can be integrated easily with this choice of $\Delta$.
\end{itemize}
To obtain the model of compact star, Dey et al. \cite{Dey2020}, Maharaj et al. \cite{Maharaj2014} choose a physically reasonable choice of $\Delta$  to find the exact solutions for the Einstein-Maxwell equations. Murad and Fatema \cite{Murad2015} obtained relativistic anisotropic charged fluid spheres by solving the
Einstein-Maxwell field equations with a preferred form of
one of the metric potentials, and suitable forms of electric
charge distribution and pressure anisotropy functions as,
\begin{eqnarray*}
V^2&=&B_N(1 + Cr^2)^N,\\
\Delta&=&\delta Cr^2(1 - 2aCr^2)(1 + Cr^2)^{1-N} \big(1 + (1 + N)Cr^2\big)^\frac{N-1}{N+1},\\
\frac{2q^2}{Cr^4}&=&K (Cr^2)^{n+1}(1+Cr^2)^{1-N} (1+mCr^2)^p\big(1+(1+N)Cr^2\big)^\frac{N-1}{N+1},
\end{eqnarray*}
where $K,\,\delta \geq 0$, $n$ is a nonnegative integer, and $m,\, p,\, a$ are
any real numbers.\\
For our present model, we assume the anisotropic factor as,
\begin{equation}\label{11}
\Delta =\frac{\left[(a + b r^2)^2-b\right]r^2}{\kappa (1 + a r^2 + b r^4)^2}.
\end{equation}
One can choose the anisotropic factor in such a way that these
allow us to integrate Eq. (\ref{fonc2}) and satisfies the physically acceptable conditions mentioned earlier. Thus this choice may be physically reasonable and useful in the study of the gravitational
behavior of anisotropic stellar objects.
By looking at the expression of (\ref{11}) and we can easily found that $\Delta(0)=0$. The above choice of $\Delta$ produces a non-negative anisotropic factor inside the stellar interior which will be discussed in the coming section. \\
Using eqns. (\ref{5}) and (\ref{11}) in (\ref{fonc2}), we get the following second order ODE in $V$ as,
\begin{eqnarray}\label{10}
V''=V'\frac{1+2ar^2+3br^4}{r(1+ar^2+br^4)},
\end{eqnarray}
In eqn. (\ref{10}), we have a second order differential equation for $V$ which on solving provides the expression of the metric coefficient for $V$ as,
\begin{eqnarray}
\label{pot2}
  V^2 &=& \left[D + \frac{C}{16 b^{\frac{3}{2}}}\times\left(2 \sqrt{b} (a + 2 b r^2)\Psi - (a^2 - 4 b) \times \right.\right.\left. \left.\log \left\{a + 2 b r^2 + 2 \sqrt{b}\Psi\right\}\right)\right]^2.
\end{eqnarray}
In the expression of $V^2$ mentioned above $C,\,D$ are constants of integration, which can be obtained from the boundary conditions.\\

With the help of (\ref{pot2}), the matter variable, radial and transverse pressure for the new solution are obtained as,
\begin{eqnarray}
\label{7}
\kappa\rho&=&\frac{3 a + (a^2 + 5 b) r^2 + 2 a b r^4 + b^2 r^6}{(1 + a r^2 + b r^4)^2},\\
\label{prsol}
  \kappa p_r &=& \frac{32 b^{3/2} C}{\Psi\big[2 \sqrt{
    b} \big(8 b D + a C \Psi+
      2 b C r^2 \Psi\big) - \Omega\big]}-\frac{a + b r^2}{\Psi^2},
      \\
			\label{ptsol}
 \kappa p_t &=&\frac{32 b^{3/2} C}{\Psi\big[2 \sqrt{
    b} \big(8 b D + a C \Psi+
      2 b C r^2 \Psi\big) - \Omega \big]}-\frac{a+2br^2}{\Psi^4}.
\end{eqnarray}
where, $\Psi,\,\Omega$ are functions of $r$ and its expression is given as, \begin{eqnarray*}\Psi=\sqrt{1 + a r^2 + b r^4},\,
\Omega=(a^2 - 4 b) C \log\left[
     a + 2 b r^2 + 2 \sqrt{b} \Psi\right].
     \end{eqnarray*}
\subsection{Physical attributes of the present model}
\subsubsection{Regularity of the metric coefficients}
To avoid the singularity, the metric potentials should take finite and positive values at the center. Now $V|_{r=0}=\frac{2 \sqrt{b} (a C + 8 b D)+(4 b-a^2) C \log{a + 2 \sqrt{b}}}{16 b^{
 3/2}}>0$ and  $W|_{r=0}=1$. We have drawn the profiles of the metric co-efficients for the compact star against $r$ in Fig~\ref{rho}. Now,
  $(V^2)'=\frac{C (r + a r^3 +
   b r^5) \big(2 \sqrt{
    b} (8 b D + a C \Psi +
      2 b C r^2 \Psi) - \Omega\big)}{8 b^{3/2} \Psi},$
    and $(W^2)'=2 r (a + 2 b r^2)$.
  We also note that $(V^2)'$ and $(W^2)'$ both vanishes at the center of the star.

    \subsubsection {Nature of the density and pressure} The central pressure ($p_c$) and central density ($\rho_c$) should be nonzero and positive valued inside the stellar interior, on the other hand $p_r$ should vanish at the boundary of the star $r=r_b$. The central pressures and the density can be written as
\begin{eqnarray}
\kappa p_c&=&\frac{32 b^{3/2} C}{
 2 \sqrt{b} (a C + 8 b D) - (a^2 - 4 b) C \log[a + 2 \sqrt{b}]}-a, \label{x1}
\\
\kappa \rho_c&=&3a,\label{x2}
\end{eqnarray}
Again from the Zeldovich's condition \cite{zel1971relativistic} $p_c/\rho_c<1$ gives the following inequality:
\begin{eqnarray}
a -  \frac{8 b^{3/2} C}{
   2 \sqrt{b} (a C + 8 b D) - (a^2 - 4 b) C \log(a + 2 \sqrt{b})}>0,\label{x3}
\end{eqnarray}
Eqns (\ref{x2}) gives, $a>0$ where as, (\ref{x1}) and (\ref{x3}) together implies,
\begin{eqnarray}
\frac{4b-a^2}{8ab}+E~<\frac{D}{C}~<\frac{16b-a^2}{8ab}+E,
\end{eqnarray}
where, $$E=\frac{a^2-4b}{16 b^{\frac{3}{2}}}\log(a + 2 \sqrt{b})$$.\\
The surface density $\rho_s$ of the compact star model is obtained as,
\begin{eqnarray}
\kappa\rho_s&=&\frac{3 a + (a^2 + 5 b) r_b^2 + 2 a b r_b^4 + b^2 r_b^6}{(1 + a r_b^2 + b r_b^4)^2}.
\end{eqnarray}
We have calculated the numerical values of the central density and surface density for different compact star in table~2.\par
\begin{figure}[htbp]
    \centering
    \includegraphics[scale=.55]{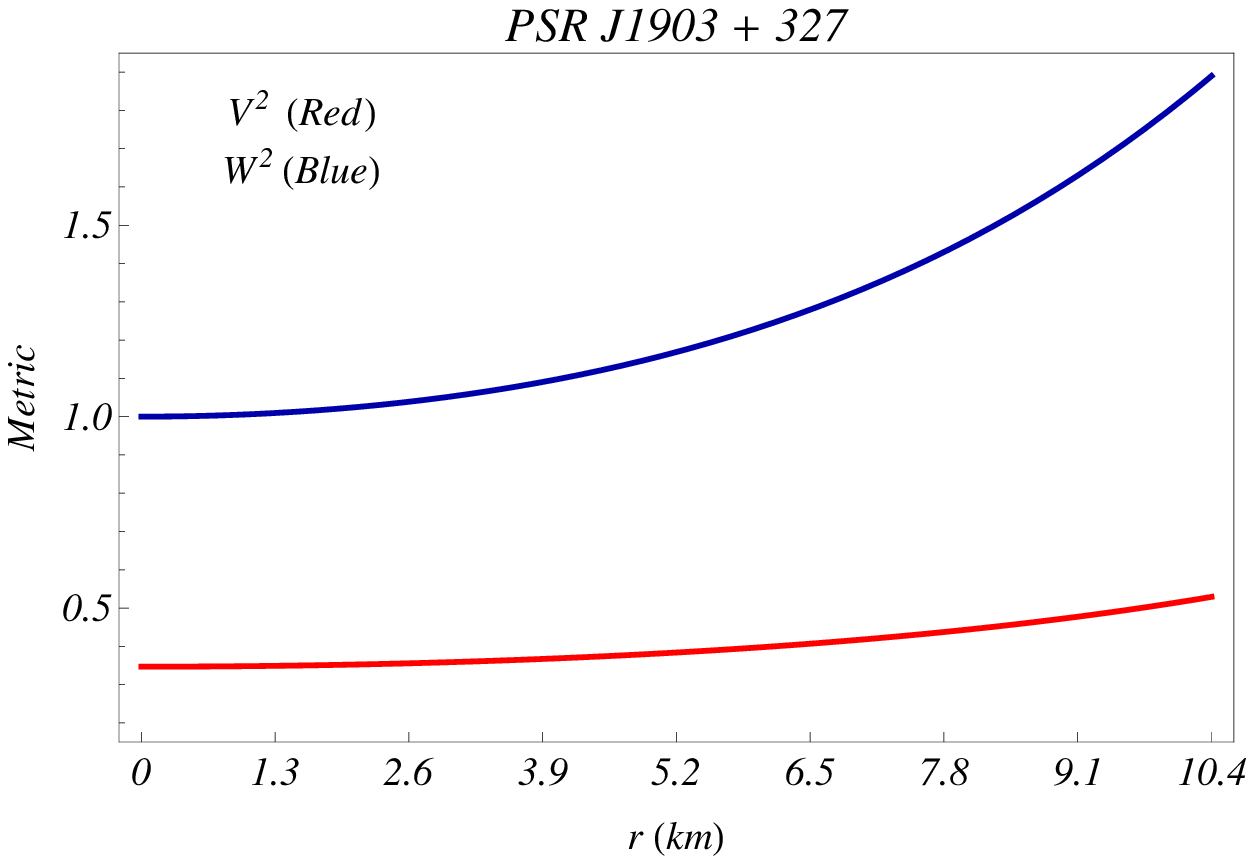}
        \includegraphics[scale=.55]{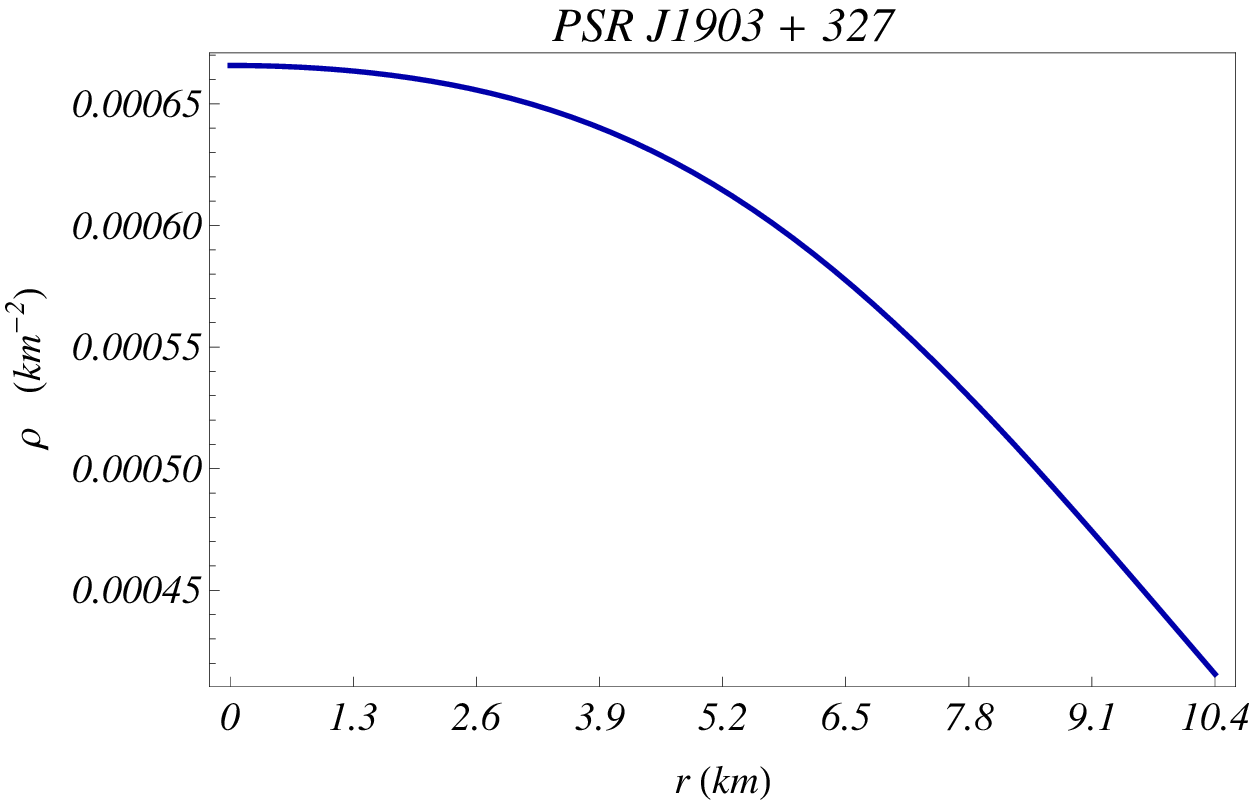}
       \caption{(Left) The metric potentials and (right) the matter density $\rho$ are plotted against $r$ inside the stellar interior for the compact star PSR J1903+327}
    \label{rho}
\end{figure}
\begin{figure}[htbp]
    \centering
        \includegraphics[scale=.55]{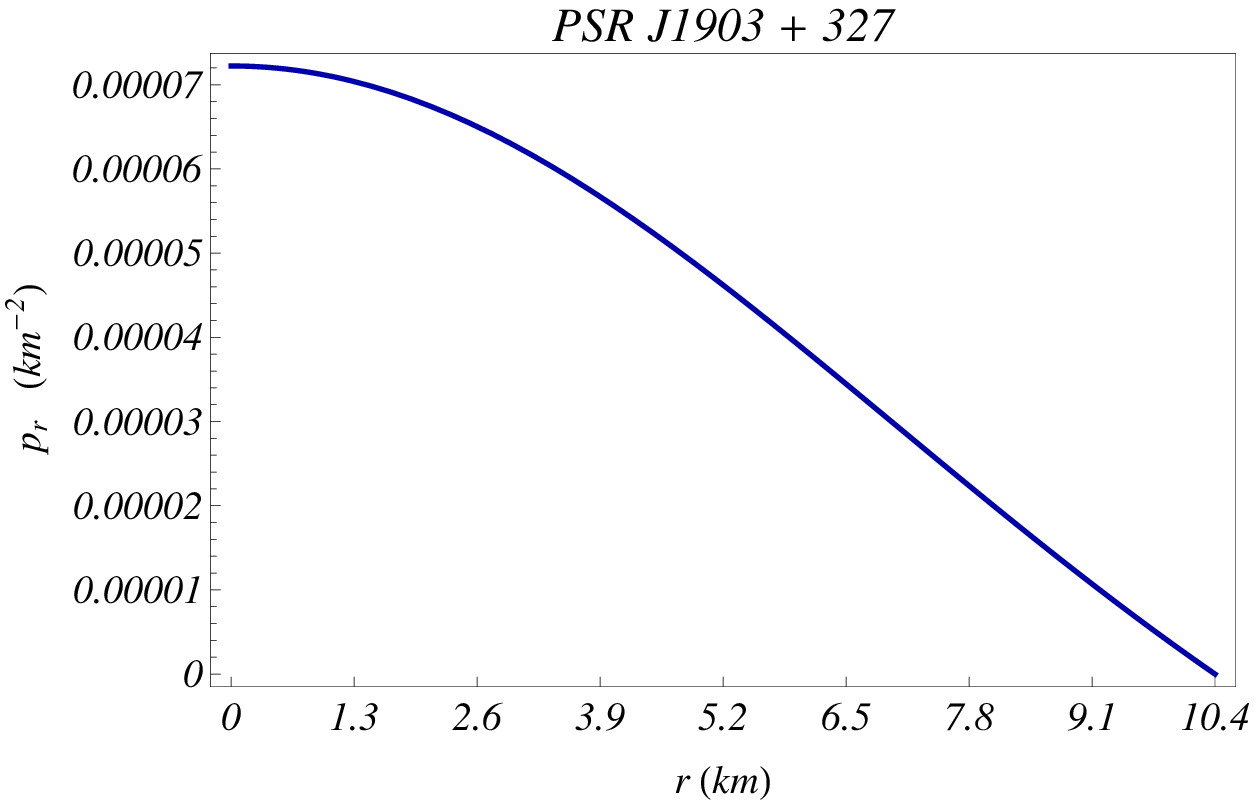}
         \includegraphics[scale=.55]{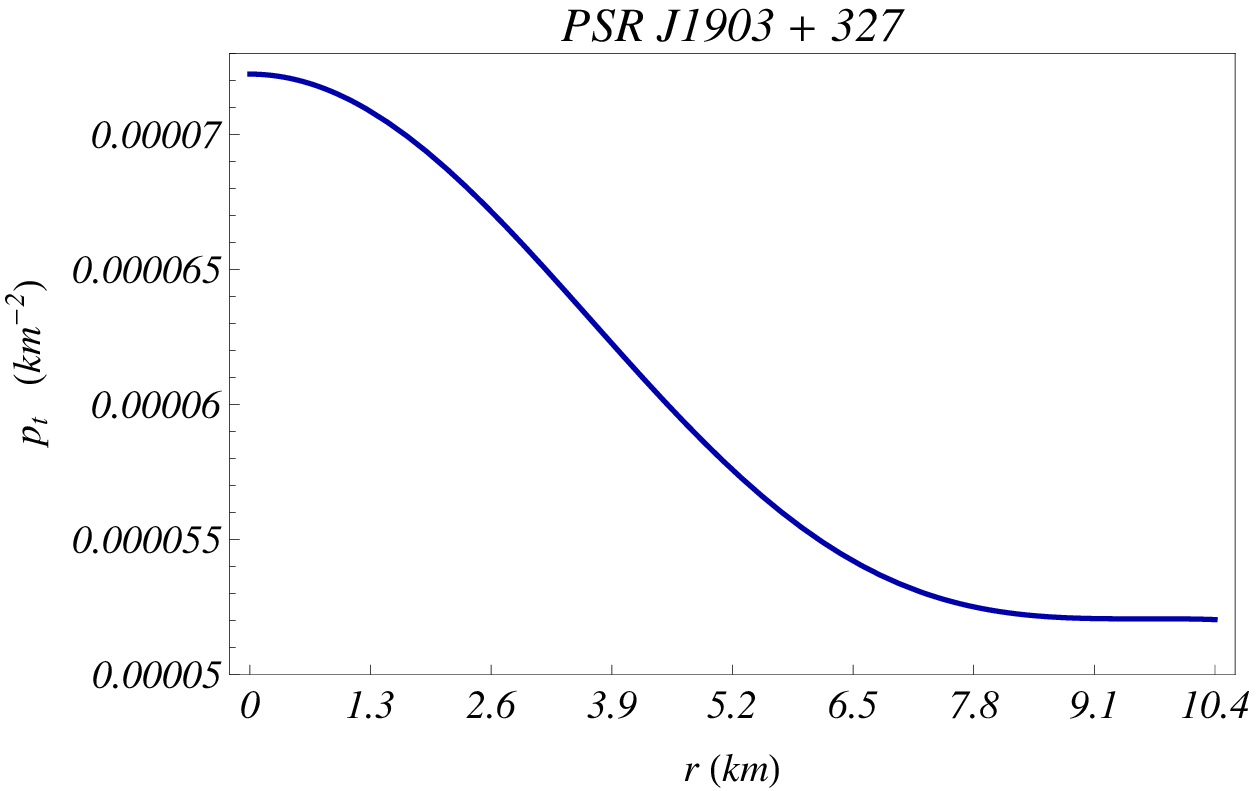}
       \caption{The radial pressure $p_r$ (left) and transverse pressure $p_t$ (right) are plotted against $r$ inside the stellar interior for the compact star PSR J1903+327 }
    \label{pr}
\end{figure}
\begin{figure}[htbp]
    \centering
        \includegraphics[scale=.55]{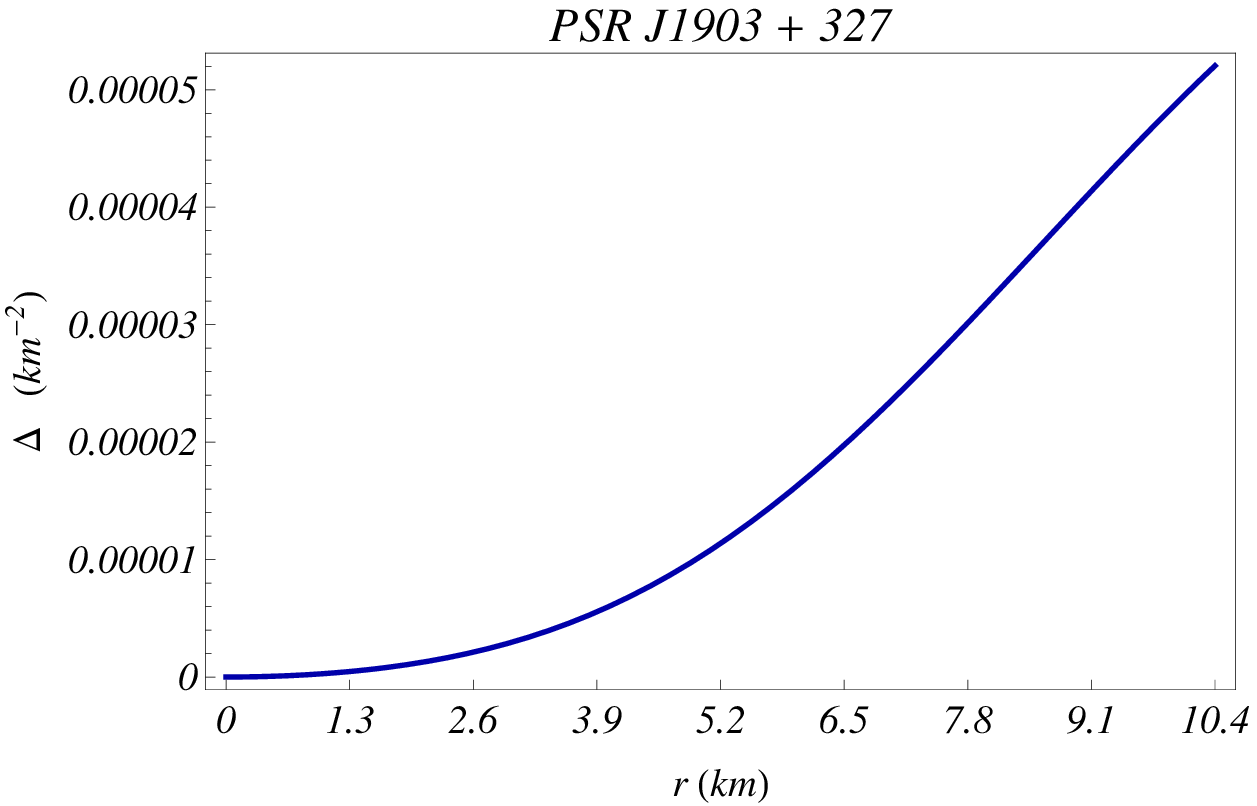}
           \includegraphics[scale=.55]{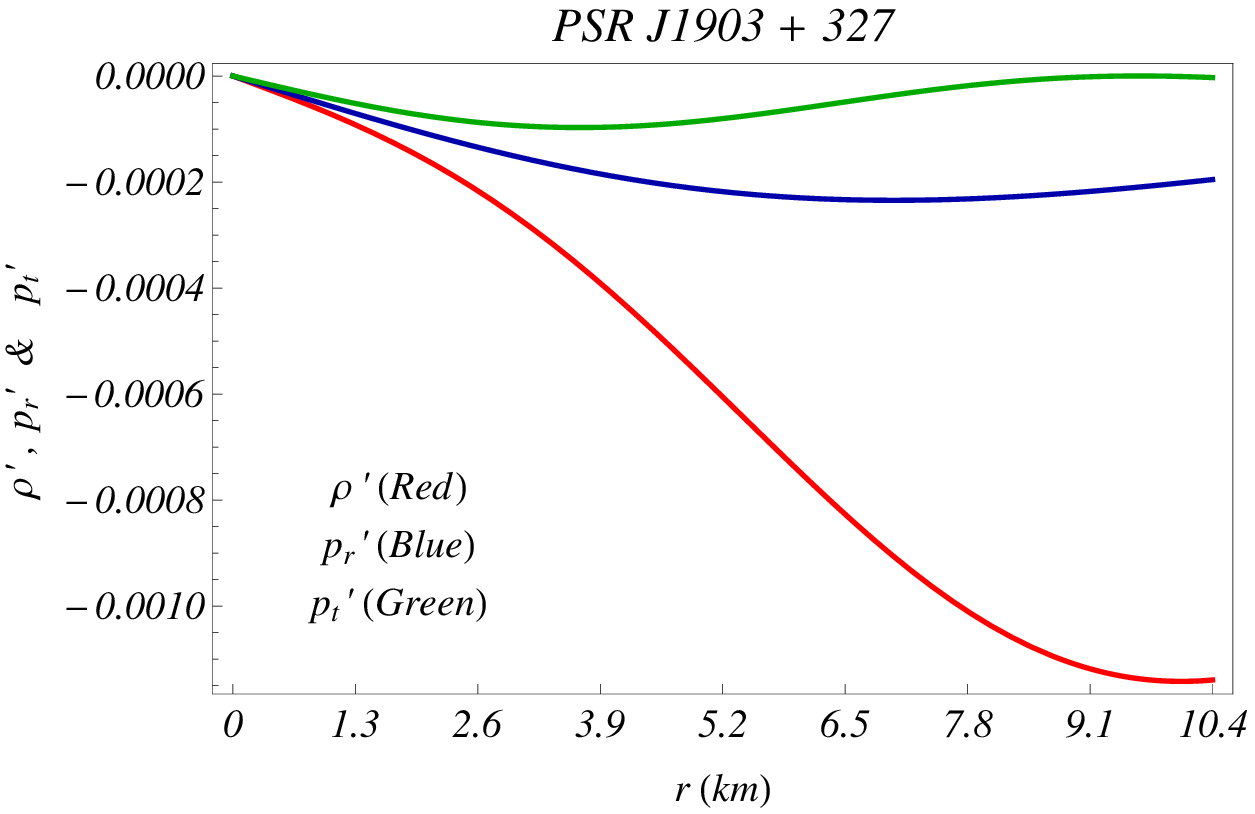}
       \caption{(Left) The anisotropic factor $\Delta$ and (right) the density and pressure gradients are plotted against $r$ inside the stellar interior for the compact star PSR J1903+327}
    \label{delta}
\end{figure}

The behavior of the matter density ($\rho$), radial and transverse pressure $p_r$ and $p_t$ inside the stellar interior are shown in Fig.~\ref{rho} and \ref{pr} respectively.\\
The density and pressure gradient of our present model are obtained by performing differentiation of the density and pressure with respect to the radial co-ordinate as,
\begin{eqnarray}
  \kappa \rho' &=& -\frac{2r}{\Psi^6}\Big[5 a^2 - 5 b + a (a^2 + 13 b) r^2 + 3 b (a^2 + 4 b) r^4 +
   3 a b^2 r^6 + b^3 r^8\Big], \\
 \kappa p_r' &=& 2 r \Big[\frac{(a + b r^2) (a + 2 b r^2)}{\Psi^4} -\frac{b}{\Psi^2}- \frac{
   16 b^{3/2} C (a + 2 b r^2)}{\Psi^3 \Phi} + \frac{
   256 b^3 C^2}{\Phi^2}\Big],\\
   \kappa p_t'&=& \kappa p_r'+\kappa \Delta',\nonumber\\
   &=&\kappa p_r'-\frac{1}{\Psi^6}\Big[2 r (-a^2 + b + a (a^2 - 5 b) r^2 + 3 (a^2 - 2 b) b r^4 +
   3 a b^2 r^6 + b^3 r^8)\Big].
\end{eqnarray}
where, \begin{eqnarray*}\Phi&=&2 \sqrt{
       b} (8 b D + a C \Psi +
         2 b C r^2 \Psi) - (a^2 - 4 b) C \log(
        a + 2 b r^2 + 2 \sqrt{b} \Psi).\end{eqnarray*}
The profiles of density and pressure gradients are shown in Fig.~\ref{delta}. We check that both the density and pressure gradients are negative and it verifies the monotonic decreasing nature of both density and pressures with respect to `r'.

The variation of anisotropy $(\Delta=p_t -p_r)$ throughout the star is regular and free of singularity.
More importantly, the anisotropy vanishes at the centre and remain positive inside the star as required.
The anisotropy profile in Fig. \ref{delta} shows that $\Delta>0$ is repulsive, allowing the construction of more compact structures.

\subsubsection{Energy Conditions}
It is well known that for a compact star model, the energy conditions should be satisfied and in this section we are interested to study about it. For an anisotropic compact star, all the energy conditions namely Weak Energy Condition (WEC), Null Energy Condition (NEC) and Strong Energy Condition (SEC)  are satisfied if and only if the following inequalities hold simultaneously for every points inside the stellar configuration.

\begin{figure}[htbp]
   \centering
        \includegraphics[scale=.55]{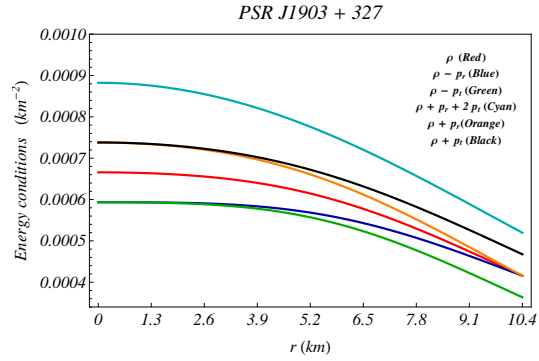}
      \caption{Energy conditions are plotted against $r$ inside the stellar interior for the compact star PSR J1903+327 }
  \label{ec}
\end{figure}

\begin{eqnarray}\label{1}
WEC &:& T_{\mu \nu}\alpha^\mu \alpha^\nu \ge 0~\Rightarrow~ \rho \geq  0,~\rho+p_r \ge 0,\,\rho+p_t \ge 0  \label{2k}\\
NEC &:& T_{\mu \nu}\beta^\mu \beta^\nu \ge 0~\Rightarrow~ \rho+p_r \geq  0,\,\rho+p_t \geq  0\\ \label{3}
DEC &:& T_{\mu \nu}\alpha^\mu \alpha^\nu \ge 0 ~\Rightarrow~ \rho \ge |p_r|,\, \rho \ge |p_t| \\ \label{4}
SEC &:& T_{\mu \nu}\alpha^\mu \alpha^\nu - {1 \over 2} T^\lambda_\lambda \alpha^\sigma \alpha_\sigma \ge 0 ~\Rightarrow~ \rho+p_r+2p_t \ge 0.\label{4k}
\end{eqnarray}
Where $~\alpha^\mu$ and $\beta^\mu$ are time-like vector and null vector respectively and $T_{\mu \nu}\alpha^{\mu} $ is nonspace-like vector. To check all the inequality stated above we have drawn the profiles of energy conditions of (\ref{2k})-(\ref{4k}) in Fig.~\ref{ec} in the interior of the compact star and this figure indicates that all the energy conditions are well satisfied by our model.

\subsubsection{Mass-Radius relation and Redshift}
The mass function of the present model can be obtained from the formula, $m(r)=\int_0^r 4\pi \rho r^2 dr$, which implies,
\begin{equation*}\label{m1}
m(r)=\frac{r^3}{2}\frac{a+br^2}{1+ar^2+br^4},
\end{equation*}
The compactness factor, i.e., the ratio of mass to the radius of a compact star for our present model is obtained from the formula,
$u=\frac{m(r)}{r}$,

\begin{figure}[htbp]
   \centering
    \includegraphics[scale=.55]{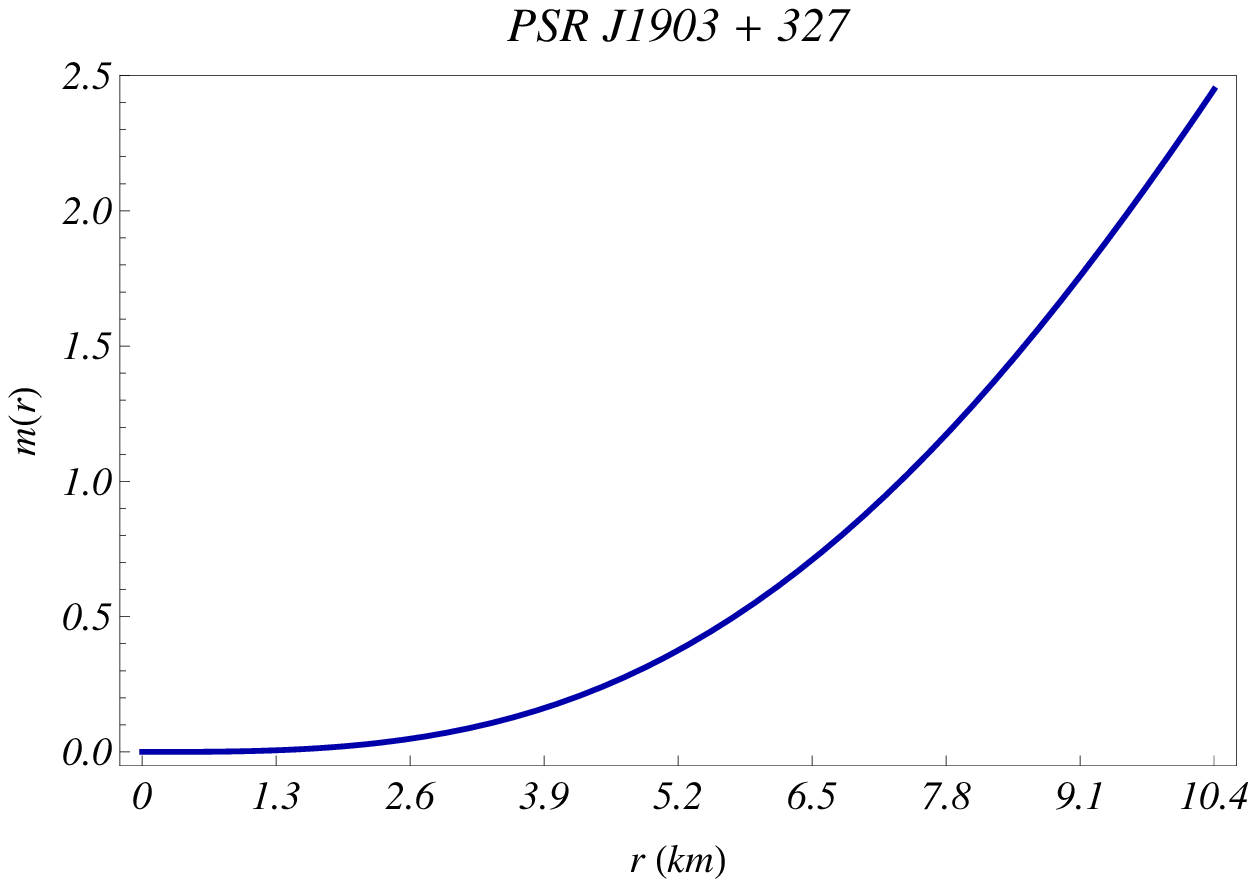}
        \includegraphics[scale=.55]{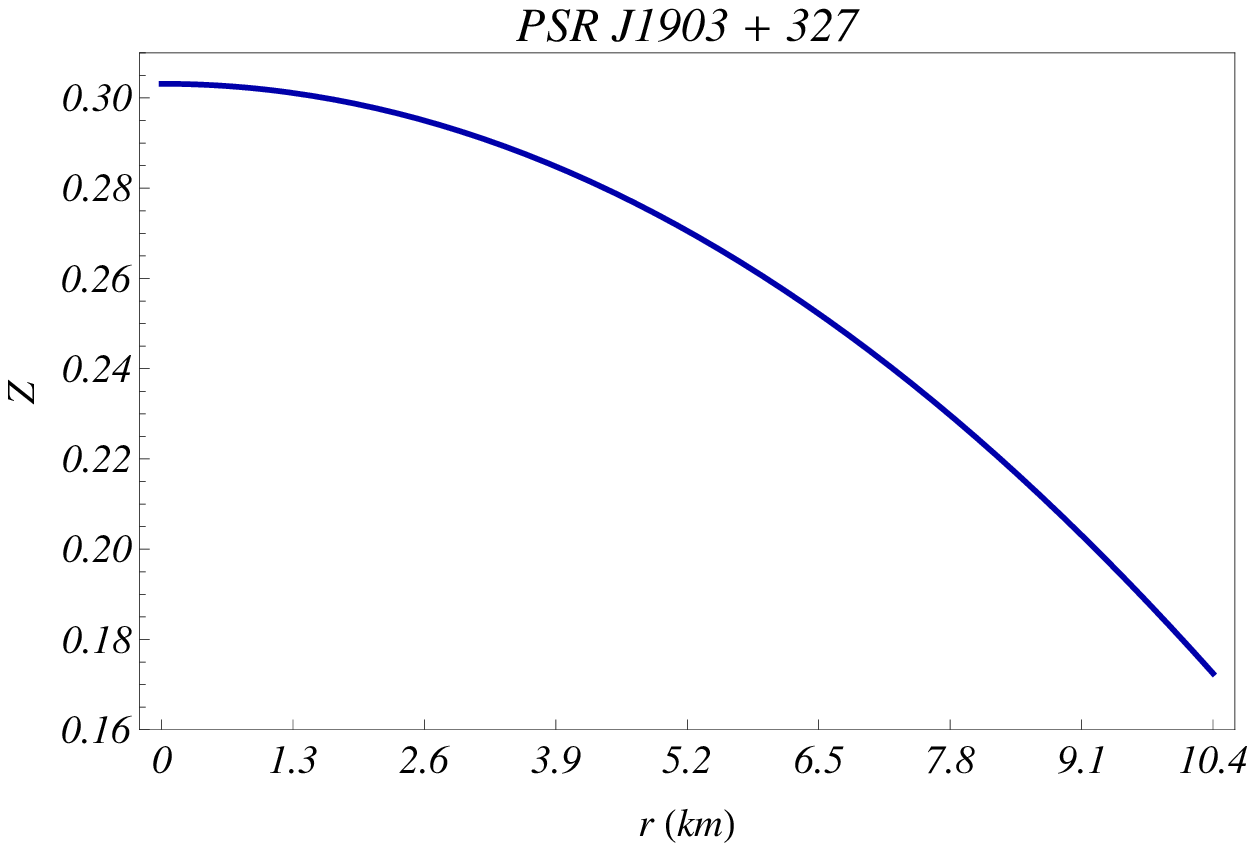}
      \caption{(Left) The mass function and (right) the gravitational redshift are plotted against $r$ inside the stellar interior for the compact star PSR J1903+327 }
  \label{grav}
\end{figure}
and consequently the surface redshift is obtained as,
\begin{eqnarray*}
z_s & = & \frac{1} {\sqrt{1-2u(r_b)}}-1=\sqrt{\frac{1+ar_b^2+br_b^4}{r_b^3(a+br_b^2)}} -1.
\end{eqnarray*}
A familiar result is that if the wavelength of a photon emitted at the surface of the star is $\lambda_e$, and the wavelength of the same
photon observed at infinity is $\lambda_o$, the gravitational red-shift $z$ is defined to be the fractional variation of the wavelength $z=\frac{\lambda_o-\lambda_e}{\lambda_e}$, which consequently gives, $z=\frac{1}{V(r)}-1$.
The central value of the gravitational redshift is obtained as,
\begin{eqnarray*}
  z_c &=&\frac{1}{V(0)}-1=\left[D + \frac{C}{16 b^{\frac{3}{2}}}\times\left\{2 \sqrt{b} a - (a^2 - 4 b) \times \right.\right.\left. \left.\log \left(a + 2 \sqrt{b}\right)\right\}\right]^{-1}-1.
\end{eqnarray*}
\section{Exterior spacetime and Boundary conditions}\label{sec3}
In this section we match our interior solution to the exterior solution smoothly at the boundary $r=r_b$ in order to find the constants $a,\,b,\,C$ and $D$. The exterior spacetime is represented by Schwarzschild vacuum solution which is given by,
\begin{eqnarray}
ds_{+}^2 & = & -f(r)dt^2+[f(r)]^{-1}dr^2+ r^2(d\theta^2+\sin^2 \theta d\phi^2),
\end{eqnarray}
corresponding to our interior line element that matches exactly with the interior solution at the boundary of the star $r=r_b$, where, $f(r)=\left(1-{2M \over r}\right)$.
For a smooth matching of the metric potentials across the boundary, i.e., at $r=r_b$,
\begin{eqnarray}\label{b1}
g_{rr}^+=g_{rr}^-,~g_{tt}^+=g_{tt}^-,
\end{eqnarray}
and
\begin{eqnarray}\label{b2}
p_r(r_b)=0.
\end{eqnarray}
Equation (\ref{b1}) gives,
\begin{eqnarray}
\sqrt{1-{2M \over r_b}}& =& D + \frac{C \left(2 \sqrt{b} (a + 2 b r_b^2)\Psi(r_b) - \Omega(r_b)\right)}{16 b^{3/2}},\label{k1}\\
  \left(1-{2M \over r_b}\right)^{-1}&=& 1+ar_b^2+br_b^4,\label{k2}
\end{eqnarray}
and (\ref{b2}) gives,
\begin{eqnarray}\label{k3}
32 b^{3/2} C\Psi(r_b)=(a + b r_b^2)\Big\{2 \sqrt{
    b} \big(8 b D + a C \Psi(r_b)+
      2 b C r_b^2 \Psi(r_b)\big)-\Omega(r_b)\Big\}
\end{eqnarray}
Solving the three equations (\ref{k1})-(\ref{k3}) the constants can be obtained as,
\begin{eqnarray}
a &=& \frac{1}{r_b^2} \left[\left(1 - \frac{2M}{r_b}\right)^{-1} - 1 - br_b^4\right],\label{z5}\\
C&=&\frac{\sqrt{\left(1 - \frac{2M}{r_b}\right)} (a + b r_b^2)}{2 \sqrt{1 + a r_b^2 + b r_b^4}},\\
D&=&\frac{C}{16 b^{
 3/2} (a + b r_b^2) \Psi(r_b)}\times \left[-2 \sqrt{
    b} (\Psi(r_b))^2 \big(a^2 + 3 a b r_b^2 +
      2 b (-8 + b r_b^4)\big)\right.\nonumber\\&&\left.+ (a^2 - 4 b)(a + b r_b^2) \Psi(r_b)
     \log\left(a + 2 b r_b^2 + 2 \sqrt{b} \Psi(r_b)\right)\right].\label{z7}
\end{eqnarray}
From eqns.~(\ref{z5})-(\ref{z7}), it is noted that if we fix the value of $b$, we obtain the values of $a,\,C$ and $D$.
The values of $a,\,C$ and $D$ for different compact stars are obtained in table~1.


\begin{table*}[t]
\centering
\caption{The values of the constants $a,\,C$ and $D$ of few well known compact star candidates.}
\label{table1}
\begin{tabular}{@{}ccccccccccc@{}}
\tableline
Objects & Observed & Observed  & Estimated &Estimated & $a$ && $C$  &&$D$ \\
&mass ($M_{\odot}$)&radius&Mass ($M_{\odot}$)& Radius &(km$^{-2}$)&&(km$^{-2}$)\\
\hline
Vela X -1 & $1.77 \pm 0.08$ & $9.56 \pm 0.08$&$1.77$&$9.5$&$0.00883866$&&$0.00232095$&&$0.738201$\\

LMC X -4  & $1.04 \pm 0.09$&$8.301 \pm 0.2$ &$1.05$&$8.1$&$ 0.00734532$&&$0.00291425$&&$0.281018$               \\
4U 1608 - 52  &$1.74 \pm 0.14$ &$9.528\pm 0.15$  &  $1.65$&$9.4$&$ 0.00825527$&&$0.00293017$&&$0.183234$   \\
PSR J1614 - 2230 &$1.97 \pm 0.04$ & $9.69 \pm 0.2$&$ 1.97$&$9.69$&$0.00975169$&&$0.00319365$&&$0.115009$       \\
EXO 1785 - 248 &$1.3 \pm 0.2$  &$8.849 \pm 0.4$   &$1.4$&$9$&  $0.00758186$&&$0.00283265$&&$0.227708$            \\
\hline

\tableline
\end{tabular}
\end{table*}


\begin{table*}[t]
\centering
\caption{The numerical values of the central density ($\rho_c$), surface density ($\rho_s$), central pressure ($p_c$), compactness factor and surface redshift of few well known compact star candidates.}
\label{table2}
\begin{tabular}{@{}cccccccccc@{}}
\tableline
Objects & $\rho_c(gm.cm^{-3})$&&$\rho_s(gm.cm^{-3})$&&$p_c(dyne.cm^{-2})$&&$\frac{2M}{R}$&&$Z_s$\\
\hline
Vela X -1& $1.07468\times10^{15}$&&$5.67945\times10^{14}$&&$2.05523\times10^{36}$&&$0.549632$&&$0.490102$\\

LMC X -4  &$1.18307\times10^{15}$&&$8.00021\times10^{14}$   &&$0.60604\times10^{35}$&&$0.382407$ &&$0.272474 $                   \\
4U 1608 - 52 &$1.32963\times10^{15}$&&$7.42043\times10^{14}$&&$1.27269\times10^{35}$ &&   $0.517819$&&$0.440108$  \\
PSR J1614 - 2230  &$1.57065\times10^{15}$ &&$7.6203\times10^{14}$  &&$2.26264\times10^{35}$ &&$0.599742$&&$0.580629$     \\
EXO 1785 - 248&$1.22116\times10^{15}$&&$7.44681\times10^{14}$&&$0.88889\times10^{35}$   &&$0.458889$&&$0.359430  $                      \\
\tableline
\end{tabular}
\end{table*}

\section{ Equation of state}\label{sec4}
Physical feature in a relativistic stellar model requires an equation of state relating the radial pressure
$p_r$ to the energy density $\rho$. Solving equation (\ref{7}) in term of $r$ and using Taylor series expansion,
we get the expression
\begin{eqnarray}
\label{eos1}
r^2&\approx &\frac{1}{\rho}+ \alpha \rho,
\end{eqnarray}
where
\begin{eqnarray}
\label{eos2}
\alpha&=& \frac{3b^2(24 \sqrt{3} a^4 b^4 -291 \sqrt{3}a^2 b^5 +1500 \sqrt{3} b^6
 + 2a^3 \beta+9ab\beta)}{\beta(2a^3b^3+9ab^4 +3\sqrt{3}\beta)},\\
 \beta&=&\sqrt{b^7(8 a^4 -97 a^2 b+500 b^2)}
\end{eqnarray}
Using the expression (\ref{eos1}) in the equation (\ref{prsol}), we get the equation of the form
\begin{eqnarray}
\label{eos}
p_r&=&-\frac{a+b\left(\frac{1}{\rho}+\alpha\rho \right)}{J^2}
+\frac{16b^3 C}{H(2\sqrt{b}(8b D+a C H+I)-J)},
\end{eqnarray}
where
\begin{eqnarray*}
H=\sqrt{1+a(\frac{1}{\rho} +\alpha \rho) +b(\frac{1}{\rho} +\alpha \rho)^2},\,
I=2b C H\left(\frac{1}{\rho}+\alpha \rho\right),\\
J=(a^2-4b) C \times\log{\left(a+\sqrt{b}H+2b\left(\frac{1}{\rho}+\alpha\rho\right) \right)}.
\end{eqnarray*}
The equation of state (\ref{eos}) is derived from the Einstein field equations, where we rewrite
the pressure in term of density. It describes a macroscopic physics features of a general relativistic gravitating compact object. It is interesting to point out, the nonlinear equation of state (\ref{eos}) contains the modified Chaplygin equation of state which has been extensively utilize in different studies. For instance, in the case of charged anisotropic fluid spheres \cite{Bhar:2016kye,Bhar:2016lfu,Bhar:2016nlq,Bhar:2015qza} and in the framework of $f(T)$ gravity theory \cite{Chanda:2019hyh,Saha:2019msh}. The Chaplygin modified equation of state has proven to be adaptable tools to investigate many open problems at the theoretical
level.

\begin{figure}[htbp]
   \centering
    \includegraphics[scale=.55]{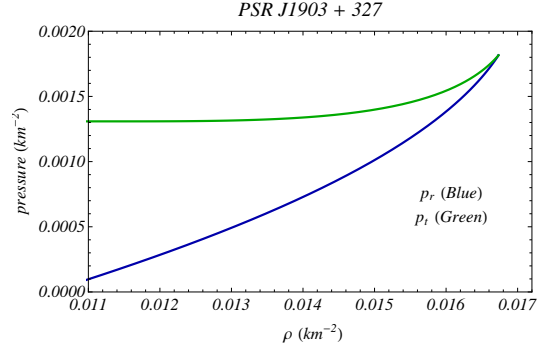}
      \caption{The variations of radial and transverse pressure with respect to density are plotted for the compact star PSR J1903+327 }
  \label{ll}
\end{figure}
The variations of radial and transverse pressure with respect to the matter density are shown in Fig.~\ref{ll}.

\section{Stability and Equilibrium condition}\label{sec5}
It is very important to check the stability of the present model. Bondi \cite{Bondi1964} developed an idea for stability analysis of neutral stars. Herrera et al. \cite{herrera92} proposed a new way for the analysis of spherical symmetric models by means of cracking (overturning), which
described the behavior of fluid distribution just after equilibrium state has been perturbed through density perturbation and this result was further extended by Gonzalez et al.\cite{gonzalez2014} by introducing local density perturbation. In this section we want to discuss the stability of the present model via (i) Harrison-Zeldovich-Novikov condition, (ii) Causality Condition and Herrera's method of cracking and (iii) Relativistic adiabatic index.
\subsection{Stability due to Harrison-Zeldovich-Novikov }
Harrison et al. \cite{Harrison1965} and Zeldovich-Novikov \cite{Novikov1972} proposed a stability condition for the model of compact star which depends on the mass and central density and it gives us information on the stability of
the gaseous stellar configuration in relation to radial pulsations. They proved that a stellar configuration will be stable if $\frac{\partial M}{\partial \rho_c}>0$, where $M=m(\rho_c)$.
For our present model,
\begin{eqnarray}
\label{partialM}
\frac{\partial M}{\partial \rho_c}=\frac{24 \pi r^3}{2(3+3\pi \rho_c r^2+3 b r^4)^2}
\end{eqnarray}
It is very clear from the above expression that $\frac{\partial M}{\partial \rho_c}>0$ and therefore the stability condition is well satisfied.

\subsection{Causality Condition and cracking}
Next we are interested to check the subliminal velocity of sound for our present model. Since we are dealing with the anisotropic fluid, the square of the radial and transverse velocity of sound $V_r^2$ and $V_t^2$ respectively should obey some bounds. According to Le Chatelier's principle, speed of sound must be positive i.e., $V_r>0,\,V_t>0$. At the same time, for anisotropic compact star model, both the radial and transverse velocity of sound should be less than $1$ which is known as causality conditions. Combining the above two inequalities one can obtain, $0<V_r^2,\,V_t^2<1$. For our present model,
\begin{eqnarray}
V_r=\sqrt{\frac{dp_r}{d\rho}}=\sqrt{\frac{p_r'}{\rho'}},\,
V_t=\sqrt{\frac{dp_t}{d\rho}}=\sqrt{\frac{p_t'}{\rho'}}
\end{eqnarray}
\begin{figure}[htbp]
    \centering
    \includegraphics[scale=.55]{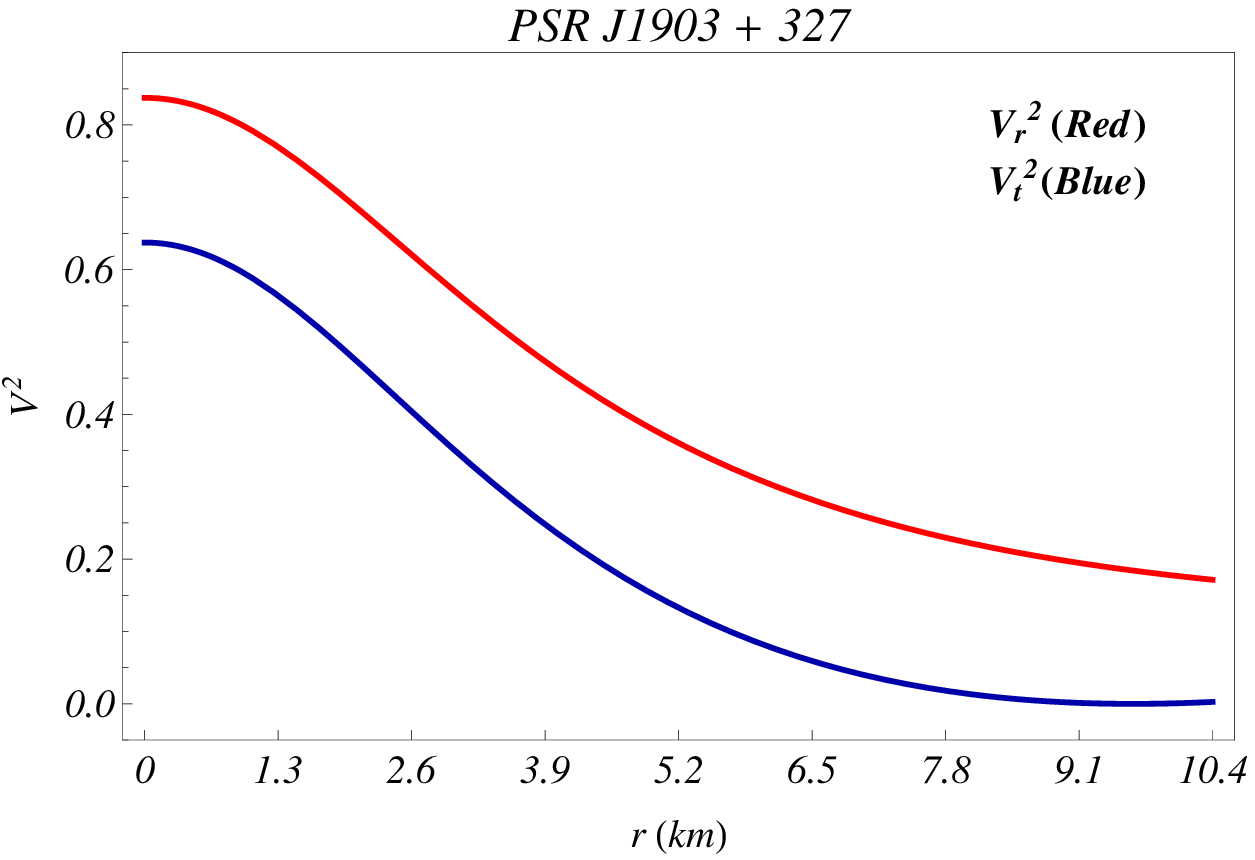}
        \includegraphics[scale=.55]{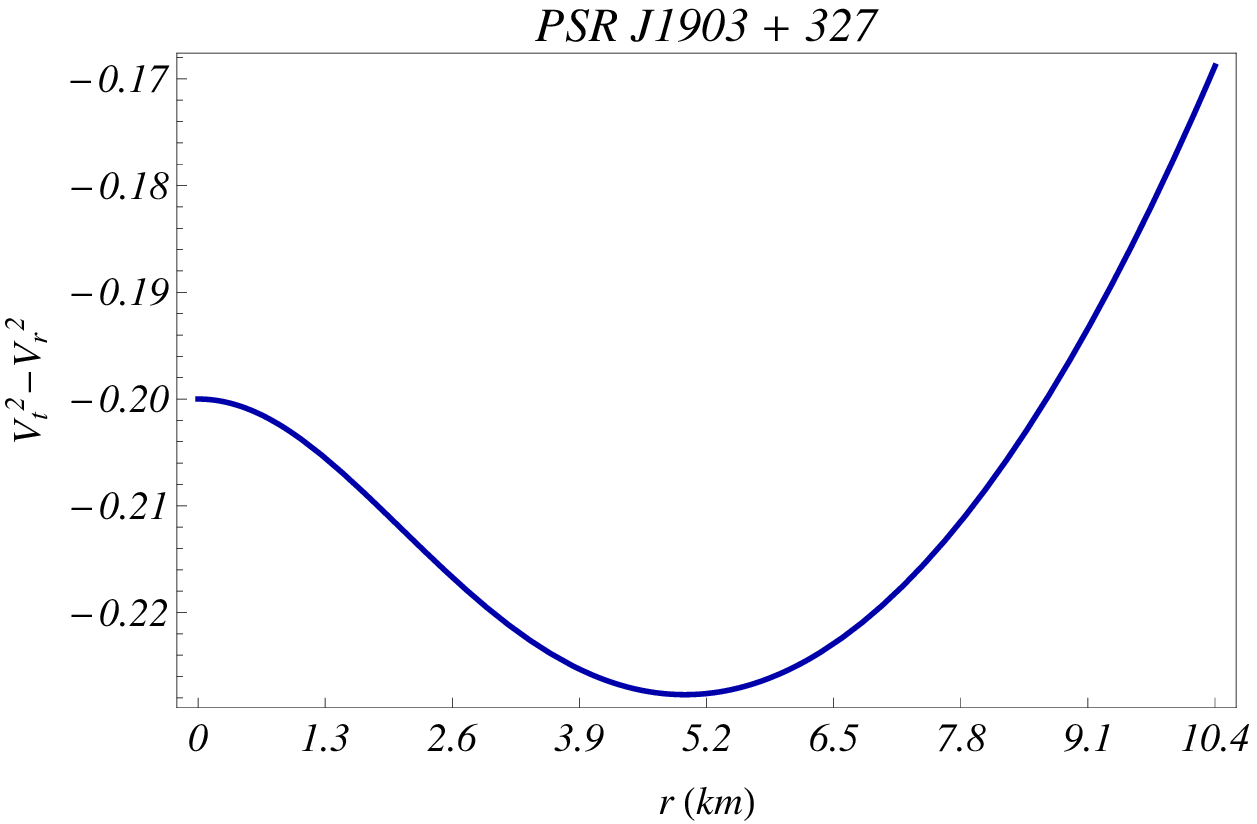}
       \caption{(Left) The variations of the square of sound velocities ($V_r^2,\,V_t^2$) and (right) the variation of $V_t^2-V_r^2$ against $r$ are shown inside the stellar interior for the compact star PSR J1903+327}
    \label{sv}
\end{figure}
The profile of radial and transverse velocity of sound are shown in Fig.~\ref{sv}.
\subsection{Relativistic Adiabatic index}
In this subsection we want to check the stability of our present model via relativistic adiabatic index. The adiabatic index $\Gamma$ is the ratio of the two specific heat and its expression can be obtained from the following formula:
\begin{eqnarray}
\Gamma&=&\frac{\rho+p_r}{p_r}V_r^2,
\end{eqnarray}

\begin{figure}[htbp]
    \centering
        \includegraphics[scale=.55]{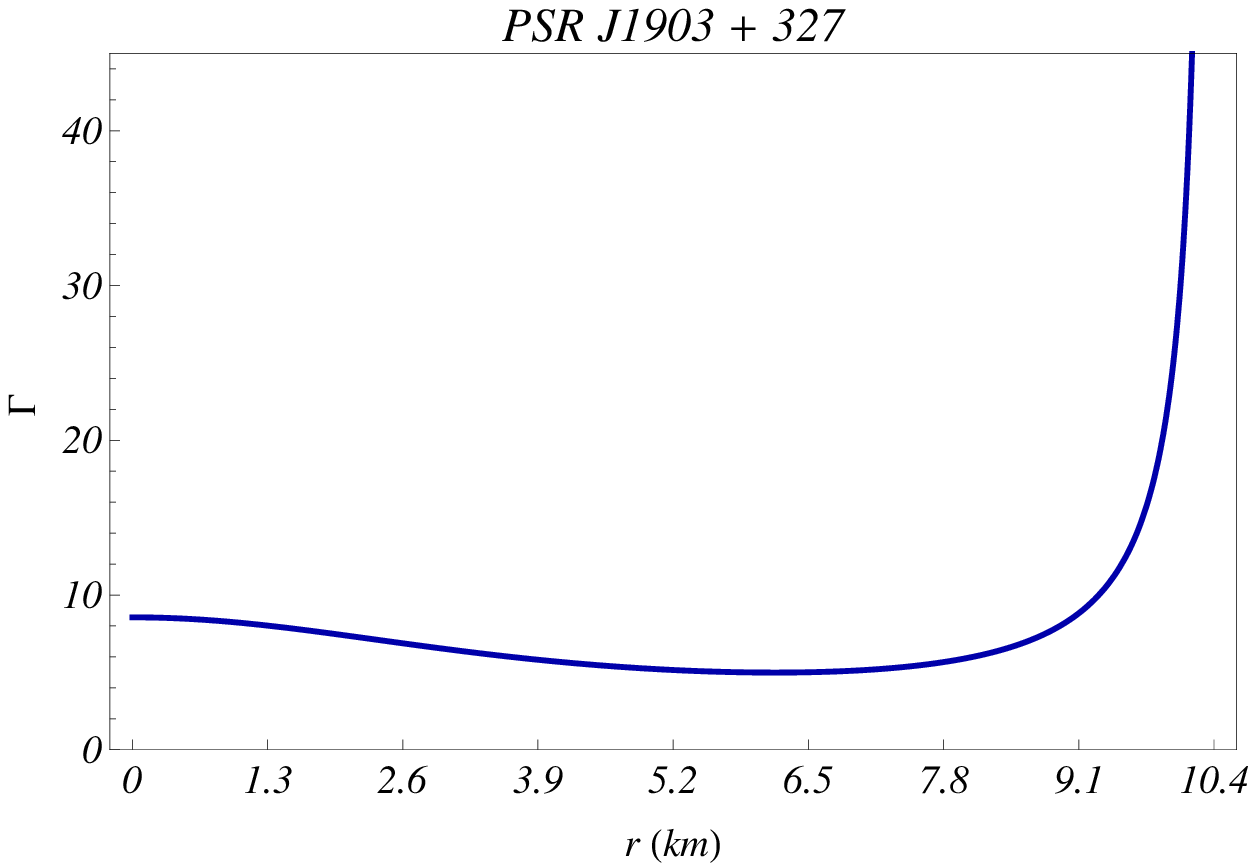}
        \includegraphics[scale=.55]{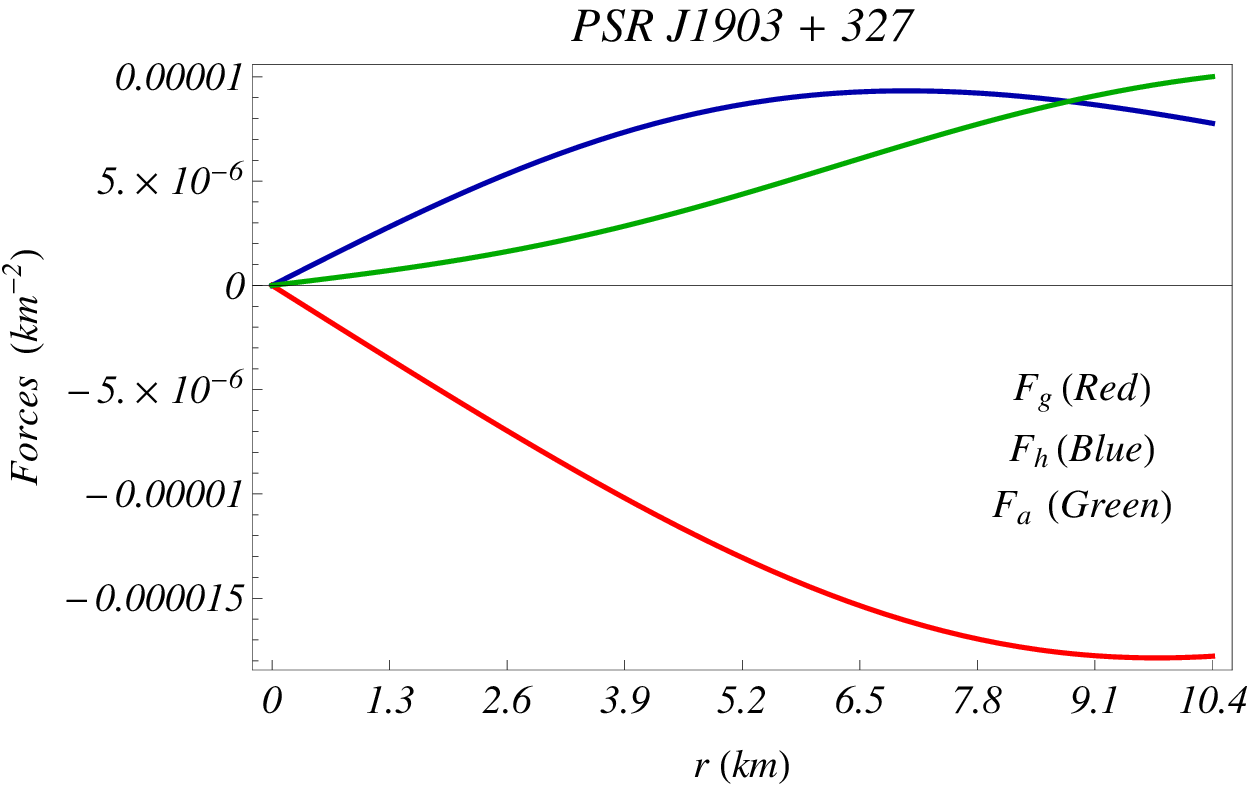}
       \caption{(Left) The adiabatic index $\Gamma$ and (right) the forces acting on the system are shown against $r$ inside the stellar interior for the compact star PSR J1903+327}
    \label{gama}
\end{figure}

Now for a newtonian isotropic sphere the stability condition is given by $\Gamma>\frac{4}{3}$ and for an anisotropic collapsing stellar configuration, the condition is quite difficult and it changes to \cite{Bondi1964}
\begin{eqnarray}
\Gamma> {4\over 3}+\left[{4\over 3}~{p_{ti}-p_{ri} \over r|p'_{ri}|}+{8\pi r \over 3}~{\rho_i p_{ri} \over |p'_{ri}|}\right]_{max} \label{gam2}
\end{eqnarray}
here $p_{ri}, ~p_{ti}$ and $\rho_i$ are initial values of radial pressure, transverse pressure and density respectively. From eqn. (\ref{gam2}), it is clear that for a stable anisotropic configuration, the limit on adiabatic index depends upon the types of anisotropy. In our present case, we have plotted the profile of $\Gamma$ and we see that it is always greater than $\frac{4}{3}$ and hence we get stable configuration from Fig. \ref{gama}.
\subsection{Equilibrium condition}
To check the static stability condition of our model under three different forces, the generalized Tolman-Oppenheimer-Volkov (TOV) equation has been considered which is represented by the equation
\begin{equation}\label{tov1}
-\frac{M_g(\rho+p_r)}{r^{2}}\frac{W}{V}-\frac{dp_r}{dr}+\frac{2}{r}(p_t-p_r)=0
\end{equation}
The above expression of $M_g(r)$ can be derived from Tolman-Whittaker mass formula and Einstein's field equation defined by
\begin{eqnarray}
M_g(r)&=&4\pi\int_0^{r}(T_t^t-T_r^r-T_{\theta}^{\theta}-T_{\phi}^{\phi})r^2 VW,
\end{eqnarray}
which further simplifies to,
\begin{equation}\label{tov2}
M_g(r)=r^{2}\frac{V'}{W},
\end{equation}. Using the expression of equation (\ref{tov2}) in (\ref{tov1}) we obtain the modified TOV equation as,
\begin{equation}
F_g+F_h+F_a=0
\end{equation}
Where the expression of the three forces are given by,
\begin{eqnarray}
F_g=-\frac{V'}{V}(\rho+p_r),\,
F_h=-\frac{dp_r}{dr},\,
F_a&=&\frac{2}{r}(p_t-p_r).
\end{eqnarray}
$F_g$, $F_h$ and $F_a$ are known as gravitational, hydro-statics and anisotropic forces respectively. The profile of the above three forces for our model of compact star is shown in Fig. \ref{gama}, which verifies that present system is in static equilibrium under these three forces.


\section{Discussion and concluding remarks}\label{sec6}
In this paper, we have presented a model of Einstein's field equations for a spherically
symmetric stellar object with anisotropic pressures. For this particular motive we have selected the Tolman-spacetime as interior geometry, and it has been traditionally matched with the Exterior Schwarzschild geometry to evaluate unknown parameters present in the model.
For the plotting of the different physical model parameters we have considered the compact star PSR $J1903+0327$ with mass $1.66~M_{\odot}$ and radius $10.4$ km \cite{fre}. From the matching conditions we obtain $a=0.00816742$, $C=0.00278563$ and $D=0.261086$.
PSR $J1903+0327$ is a millisecond pulsar in a highly eccentric binary orbit and it lies in the categories of stars whose masses have
been found accurately \cite{fre}. The pulse period is $2.15$ ms, or $465.1$ times per second. These stars are composed of the densest material exist in this universe. The radii
of these stars depend upon the equation of state (EoS) i.e.,
how physical variables are related to each other \cite{azam}. Following points summarize our concluding remarks :
\begin{itemize}
\item As, if one analyze the results from left panel of Fig.~\ref{rho}, it is well clear that both the metric gravitational potentials have positive and regular behavior inside the stellar interior. The study of compact stars requires the metric gravitational components must behave positively (i.e. $V,\,W>0$) when graphes against radial coordinate `r'.
\item Energy progression as shown in the right panel of Fig.~\ref{rho} promises the real formation of stellar body by having positive advancement all over the matter distribution along radial direction and shows smooth decline from inward to outward direction.
\item    Congruent to density, the behavior of pressure components in left and right panel of the Fig.~\ref{pr} declares the realistic formation of compact object as one can assess that $p_t>0$ and $p_r$ approaches to zero exactly at the boundary $r=R$. Both the pressures are positive, continuous and do not suffer from any kind of singularities inside the stellar interior.
\item     Out spread of anisotropic factor ($\Delta\geq0$) Shown in the left panel of Fig.~\ref{delta} ensures the existence of repulsive forces allowing the more compact formation by escaping the stellar collapsing to point singularity. Right panel in the Fig. \ref{delta} depicts the negative trend (Propagating from zero to negative from center to boundary) in gradient components which strengthen the equilibrium of the system by counterbalancing the forces inside the stellar body.
    \item    The model analysis shows that the null, weak, dominant and strong energy conditions are satisfied throughout the stellar structure as shown in Fig. \ref{ec}. For the complexity of the expressions of density and pressures we have taken the help of graphical representations which ensures about the well behaved nature of all the energy conditions.

    \item The mass function and gravitational redshift are plotted in Fig.~\ref{grav}. The figure shows that mass function is monotonic increasing function of `r' and regular at the center of the star. On the other hand the gravitational redshift is monotonic decreasing. The mass and gravitational redshift values are in agreement with required physical conditions as one can examine from Fig \ref{grav}.
For a spherical object, in the absence of a cosmological constant,
Buchdahl \cite{Buchdahl1959} has proposed the upper bound for surface redshift as,
$z_s \leq 2$ which was generalized by B\"{o}hmer and Harko \cite{bohmer2007minimum} for an
anisotropic spherical object in the presence of a cosmological constant $\Lambda$ as
$z_s \leq 5$. Later this bound was modified by Ivanov \cite{Ivanov2002}
who demonstrated that the most extreme admissible value
could be as high as $z_s = 5.211$. In our present study we have obtained the surface redshift for different compact star models presented in table~\ref{table2}, which indicates that $z_s$ lies in the reasonable bound.
\item     The variation of radial and transverse pressures with respect to the density has been pointed in Fig. \ref{ll}. The figure indicates that both the pressures obeys a non-linear equation of state with respect to matter density.
\item Sound speeds and Herrera's cracking conditions are plotted in Fig. \ref{sv} versus radial direction, both the sound speeds are within the stable range of compact star i.e. $0<V_r^2,\,V_t^2<1$. Herrera's cracking condition \cite{herrera92} $V^2_{t}-V^2_{r}<0$ is also justified to ensure the potentially stability of the system.
\item     As TOV forces are also in good balance to declare the equilibrium state of the stellar model of our case study as shown in right panel of Fig. \ref{gama}. Left panel of Fig. \ref{gama} shows the stability of the system of stellar objects under adiabatic index $\Gamma>\frac{4}{3}$.

\end{itemize}
The model is potentially stable and regular and more details numerical features of our solution can be found in Table \ref{table1} and \ref{table2}.
The masses and radii of some pulsars obtained from our model are roughly equal to the observed stars such
Vela X -1, LMC X -4, 4U 1608 - 52, PSR J1614 - 2230, EXO 1785 - 248.
Through analytical, graphical and numerical analysis, all the features of the model are well described. Finally, if we summarize our discussion, we collectively convinced by the calculated results which says that the system under discussion is physically admissible and viably stable.\\

\section*{ACKNOWLEDGMENTS} PB is thankful to IUCAA, Government of India for providing visiting associateship.
PMT thanks the University of South Africa and National Research Foundation for financial support.

\bibliography{epjc2}

\end{document}